\documentclass[3p,twocolumn,sort&compress]{elsarticle}

    \makeatletter
    \def\ps@pprintTitle{%
      \let\@oddhead\@empty
      \let\@evenhead\@empty
      \def\@oddfoot{\reset@font\hfil\thepage\hfil}
      \let\@evenfoot\@oddfoot
    }
    \makeatother
    
\usepackage{graphicx}
\usepackage{bm}
\usepackage{wasysym}

\newcommand\ee{\end{equation}}
\newcommand\be{\begin{equation}}
\newcommand\eea{\end{eqnarray}}
\newcommand\bea{\begin{eqnarray}}

\begin{document}

\title{{\bf A new approach to evaluate the leading \\ hadronic corrections to the muon \boldmath $g$-2\unboldmath}~\tnoteref{t1}}\tnotetext[t1]{This work is dedicated to the memory of our friend and colleague Eduard A.\ Kuraev.}

\author[carlo]{C.\ M.\ Carloni Calame}
\ead{carlo.carloni.calame@pv.infn.it}
\author[massimo]{M.\ Passera}
\ead{massimo.passera@pd.infn.it}
\author[luca]{L.\ Trentadue}
\ead{luca.trentadue@cern.ch}
\author[graziano]{G.\ Venanzoni}
\ead{graziano.venanzoni@lnf.infn.it}

\address[carlo]{Dipartimento di Fisica, Universit\`a di Pavia, Pavia, Italy}
\address[massimo]{INFN, Sezione di Padova, Padova, Italy}
\address[luca]{Dipartimento di Fisica e Scienze della Terra 
``M.\ Melloni''\\Universit\`a di Parma, Parma, Italy and \\ INFN, Sezione di Milano Bicocca, Milano, Italy}
\address[graziano]{INFN, Laboratori Nazionali di Frascati, Frascati, Italy}

\begin{abstract}
We propose a novel approach to determine the leading hadronic corrections to the muon $g$-2. It consists in a measurement of the effective electromagnetic coupling in the space-like region extracted from Bhabha scattering data. We argue that this new method may become feasible at flavor factories, resulting in an alternative determination potentially competitive with the accuracy of the present results obtained with the dispersive approach via time-like data.
\end{abstract}
\maketitle

\section{Introduction}

The long-standing discrepancy between experiment and the Standard Model (SM) prediction of $a_{\mu}$, the muon anomalous magnetic moment, has kept the hadronic corrections under close scrutiny for several years~\cite{Bennett:2006fi,Jegerlehner:2009ry,Reviews,Jegerlehner:2008zza}. In fact, the hadronic uncertainty dominates that of the SM value and is comparable with the experimental one. When the new results from the $g$-2 experiments at Fermilab and J-PARC will reach the unprecedented precision of 0.14 parts per million (or better)~\cite{Grange:2015fou,Venanzoni:2014ixa,Saito:2012zz}, the uncertainty of the hadronic corrections will become the main limitation of this formidable test of the SM.

An intense research program is under way to improve the evaluation of the leading order (LO) hadronic contribution to $a_{\mu}$, due to the hadronic vacuum polarization correction to the one-loop diagram~\cite{Venanzoni:2014wva,Fedotovich:2008zz}, as well as the next-to-leading order (NLO) hadronic one. The latter is further divided into the $O(\alpha^3)$ contribution of diagrams containing hadronic vacuum polarization insertions~\cite{Krause:1996rf}, and the leading hadronic light-by-light term, also of $O(\alpha^3)$~\cite{Jegerlehner:2009ry,HLBL,HLBLfuture}. Very recently, even the next-to-next-to leading order (NNLO) hadronic contributions have been studied: insertions of hadronic vacuum polarizations were computed in~\cite{Kurz:2014wya}, while hadronic light-by-light corrections have been estimated in~\cite{Colangelo:2014qya}.

The evaluation of the hadronic LO contribution $a_{\mu}^{\scriptscriptstyle \rm HLO}$ involves long-distance QCD for which perturbation theory cannot be employed. However, using analyticity and unitarity, it was shown long ago that this term can be computed via a dispersion integral using the cross section for low-energy hadronic $e^+ e^-$ annihilation~\cite{BM61GDR69}. At low energy this cross-section is highly fluctuating due to resonances and particle production threshold effects.

As we will show in this paper, an alternative determination of $a_{\mu}^{\scriptscriptstyle \rm HLO}$ can be obtained measuring the effective electromagnetic coupling in the space-like region extracted from Bhabha ($e^+ e^- \to e^+ e^-$) 
scattering data. A method to determine the running of the electromagnetic coupling in small-angle Bhabha scattering was proposed in~\cite{Arbuzov:2004wp} and applied to LEP data in~\cite{Abbiendi:2005rx}. As vacuum polarization in the space-like region is a smooth function of the squared momentum transfer, the accuracy of its determination is only limited by the statistics and by the control of the systematics of the experiment. Also, as at flavor factories the Bhabha cross section is strongly enhanced in the forward region, we will argue that a space-like determination of $a_{\mu}^{\scriptscriptstyle \rm HLO}$ may not be limited by statistics and, although challenging, may become competitive with standard results obtained with the dispersive approach via time-like data.

\section{Theoretical framework}

The leading-order hadronic contribution to the muon $g$-2 is given by the well-known formula~\cite{BM61GDR69,Jegerlehner:2008zza}
\begin{equation}\label{amu}
	a_{\mu}^{\scriptscriptstyle \rm HLO} = \frac{\alpha}{\pi^2} \int_0^{\infty} \frac{ds}{s} 
	\;K(s)\; {\rm Im} \Pi_{\rm had}(s+i \epsilon),
\end{equation}
where $\Pi_{\rm had}(s)$ is the hadronic part of the photon vacuum polarization, 
$\epsilon>0$, 
\be
	K(s) = \int_0^1 dx \, \frac{x^2 (1-x)}{x^2 + (1-x) (s/m_{\mu}^2)}
\ee
is a positive kernel function, and $m_{\mu}$ is the muon mass. As the total cross section for hadron production in low-energy $e^+ e^-$ annihilations is related to the imaginary part of  $\Pi_{\rm had}(s)$ via the optical theorem, the dispersion integral in eq.~(\ref{amu}) is computed integrating experimental time-like ($s>0$) data up to a certain value of $s$~\cite{Jegerlehner:2009ry, Davier:2010nc, Hagiwara:2011af}. The high-energy tail of the integral is calculated using perturbative QCD~\cite{pQCD}.

Alternatively, if we exchange the $x$ and $s$ integrations in eq.~(\ref{amu}) we obtain~\cite{Lautrup:1971jf}
\begin{equation}
	a_{\mu}^{\scriptscriptstyle \rm HLO} = 
	 \frac{\alpha}{\pi} \int_0^1 dx \, (x-1) \,  \overline{\Pi}_{\rm had} \!  \left[ t(x) \right],
\label{amu_x}
\end{equation}
where $\overline{\Pi}_{\rm had}(t) = \Pi_{\rm had}(t) - \Pi_{\rm had}(0)$ and 
\begin{equation}
	t(x)=\frac{x^2m_\mu^2}{x-1}  < 0
\label{t}
\end{equation}
is a space-like squared four-momentum. If we invert eq.~(\ref{t}), we get
$
	x=\left( 1- \beta \right) (t/2m_{\mu}^2),
$
with 
$\beta = (1-4m_{\mu}^2/t)^{1/2}$, and from eq.~(\ref{amu_x}) we obtain
\begin{equation}\label{amu_t}
	a_{\mu}^{\scriptscriptstyle \rm HLO} = 
	\frac{\alpha}{\pi} \int_{-\infty}^0  \overline{\Pi}_{\rm had} (t)
	\left( \frac{\beta -1}{\beta+1} \right)^2
	\frac{dt}{t \beta}.
\end{equation}
Equation (\ref{amu_t}) has been used for lattice QCD calculations of $a_{\mu}^{\scriptscriptstyle \rm HLO}$~\cite{lattice}; while the results are not yet competitive with those obtained with the dispersive approach via time-like data, their errors are expected to decrease significantly in the next few years~\cite{Blum:2013qu}.

The effective fine-structure constant at squared momentum transfer $q^2$ can be defined by
\begin{equation}\label{alphaq2}
        \alpha(q^2) = \frac{\alpha}{1-\Delta \alpha(q^2)},
\end{equation}
where
$
 	\Delta \alpha(q^2) = - {\rm Re} \overline{\Pi} (q^2).
$
The purely leptonic part, $\Delta \alpha_{\rm lep} (q^2)$, can be calculated order-by-order in perturbation theory -- it is known up to three loops in QED~\cite{Steinhauser:1998rq} (and up to four loops in specific $q^2$ limits~\cite{4loop}). As Im$\overline{\Pi}(q^2)=0$ for negative $q^2$, eq.~(\ref{amu_x}) can be rewritten in the form~\cite{Jegerlehner:2001wq}
\begin{equation}\label{amu_xalpha}
	a_{\mu}^{\scriptscriptstyle \rm HLO} = 
	 \frac{\alpha}{\pi} \int_0^1 dx \, (1-x) \,  \Delta \alpha_{\rm had} \! \left[ t(x) \right].
\end{equation}
Equation (\ref{amu_xalpha}), involving the hadronic contribution to the running of the effective fine-structure constant at space-like momenta, can be further formulated in terms of the Adler function~\cite{Adler:1974gd}, defined as the logarithmic derivative of the vacuum polarization, which, in turn, can be calculated via a dispersion relation with time-like hadroproduction data and perturbative QCD~\cite{Eidelman:1998vc,Jegerlehner:2001wq}. We will proceed differently, proposing to calculate eq.~(\ref{amu_xalpha}) by measurements of the effective electromagnetic coupling in the space-like region (see also~\cite{Fedotovich:2008zz}).

\section{$\Delta\alpha_{\rm had}(t)$ from Bhabha scattering data}

The hadronic contribution to the running of $\alpha$ in the space-like region, $\Delta\alpha_{\rm had}(t)$, can be extracted comparing Bhabha scattering data to Monte Carlo (MC) predictions. The LO Bhabha cross section receives contributions from $t$- and $s$-channel photon exchange amplitudes. At NLO in QED, it is customary to distinguish corrections with an additional virtual photon or the emission of a real photon (photonic NLO) from those originated by the insertion of the vacuum polarization corrections into the LO photon propagator (VP). The latter goes formally beyond NLO when the Dyson resummed photon propagator is employed, which simply amounts to rescaling the $\alpha$ coupling in the LO $s$- and $t$-diagrams by the factor $1/(1-\Delta\alpha(q^2))$ (see eq.~(\ref{alphaq2})). In MC codes, e.g.\ in \texttt{BabaYaga}~\cite{babayaga}, VP corrections are also applied to photonic NLO diagrams, in order to account for a large part of the effect due to VP insertions in the NLO contributions. Beyond NLO accuracy, MC generators consistently include also the exponentiation of (leading-log) QED corrections to provide a more realistic simulation of the process and to improve the theoretical accuracy. We refer the reader to ref.~\cite{Actis:2010gg} for an overview of the status of the most recent MC generators employed at flavor factories. We stress that, given the inclusive nature of the measurements, any contribution to vacuum polarization which is not explicitly subtracted by the MC generator will be part of the extracted $\Delta\alpha(q^2)$. This could be the case, for example, of the contribution of hadronic states including photons (which, although of higher order, are conventionally included in $a_{\mu}^{\scriptscriptstyle \rm HLO}$), and that of $W$ bosons or top quark pairs.

\begin{figure*}[htp]
  \includegraphics[width=.5\textwidth]{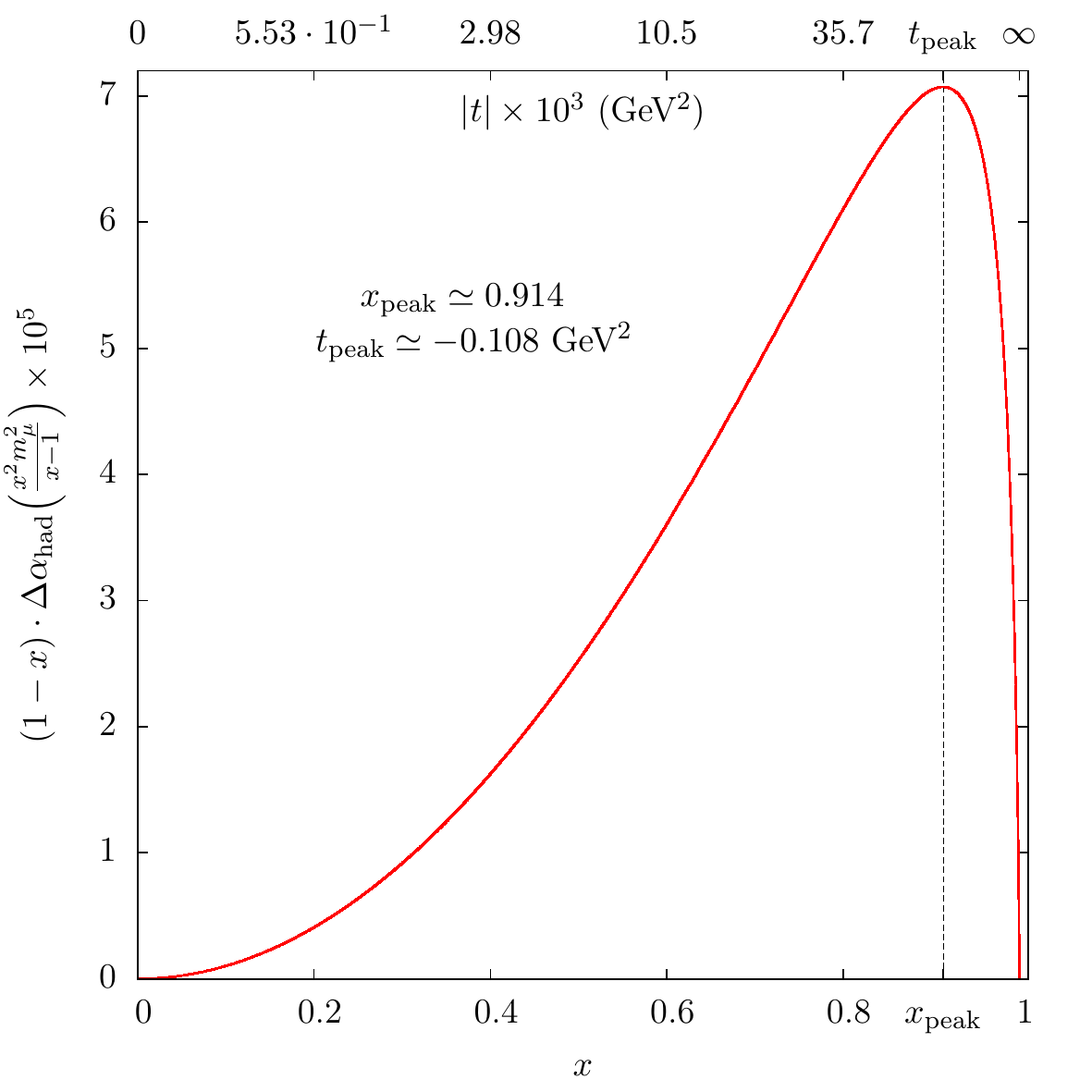}~\includegraphics[width=.5\textwidth]{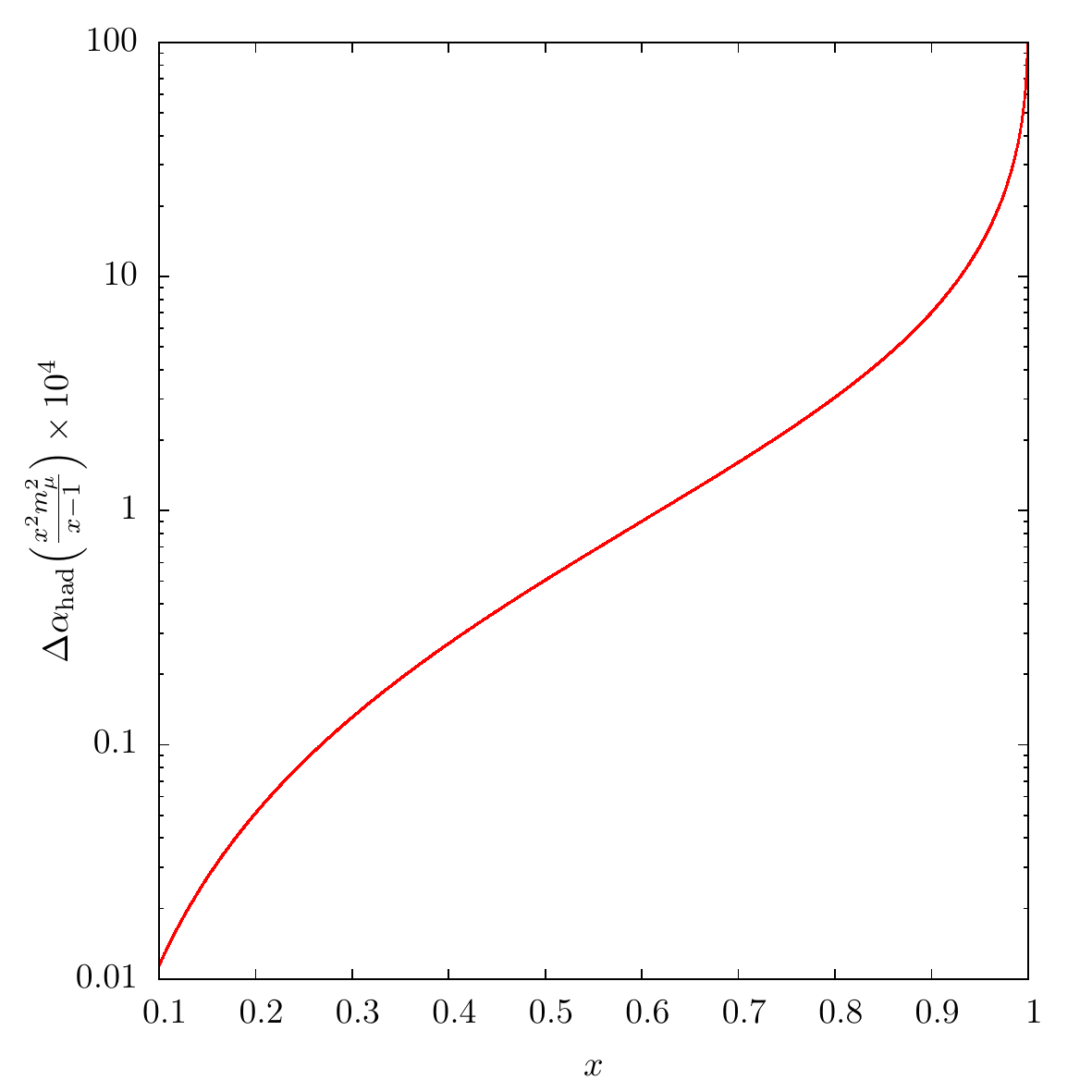}
  \caption{Left: The integrand $(1-x)\Delta\alpha_{\rm had}[t(x)] \times 10^5$ as a function of $x$ and $t$. Right: $\Delta\alpha_{\rm had}[t(x)] \times 10^4$.}
  \label{xfunction}
\end{figure*}

\begin{figure*}[htp]
  \includegraphics[width=.5\textwidth]{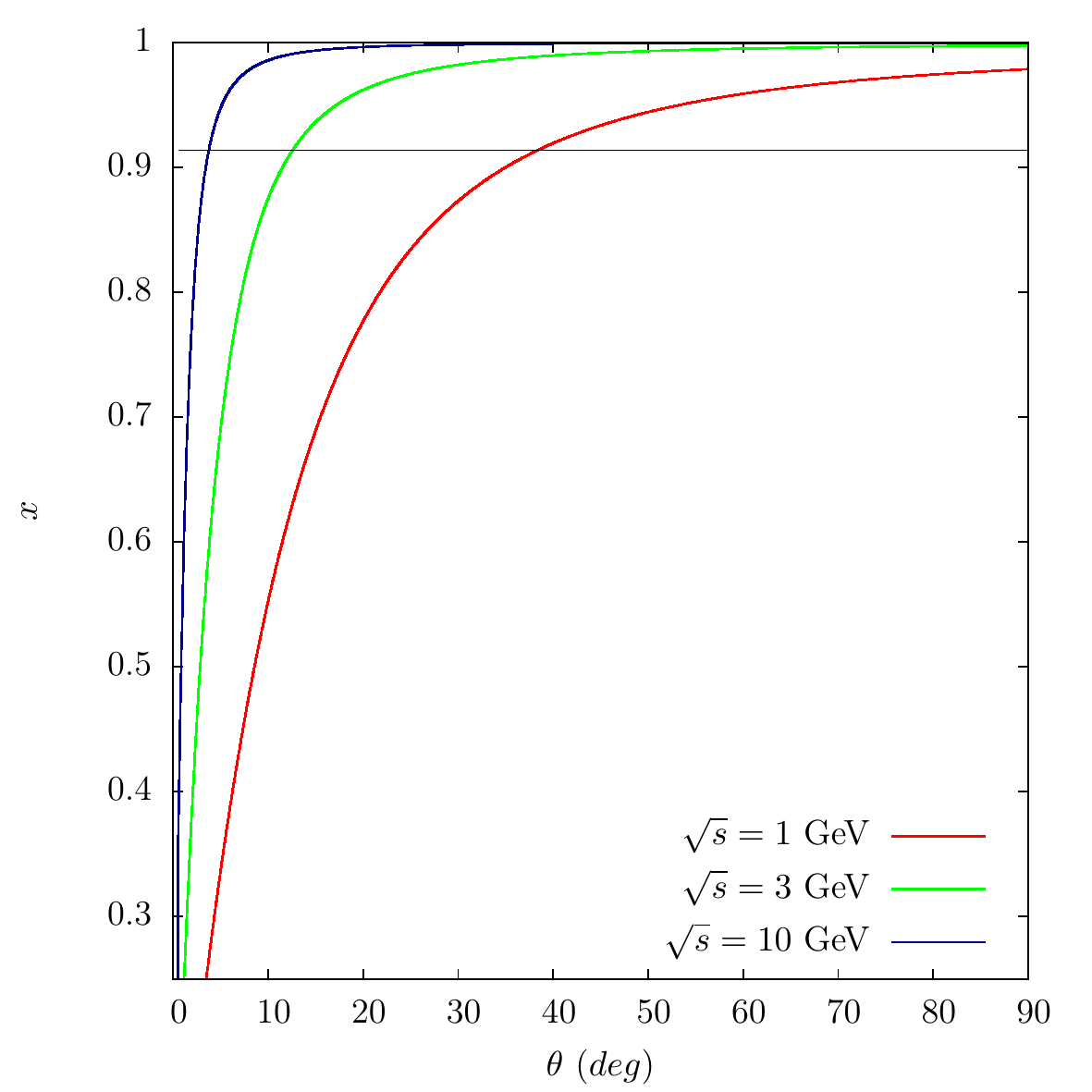}~\includegraphics[width=.5\textwidth]{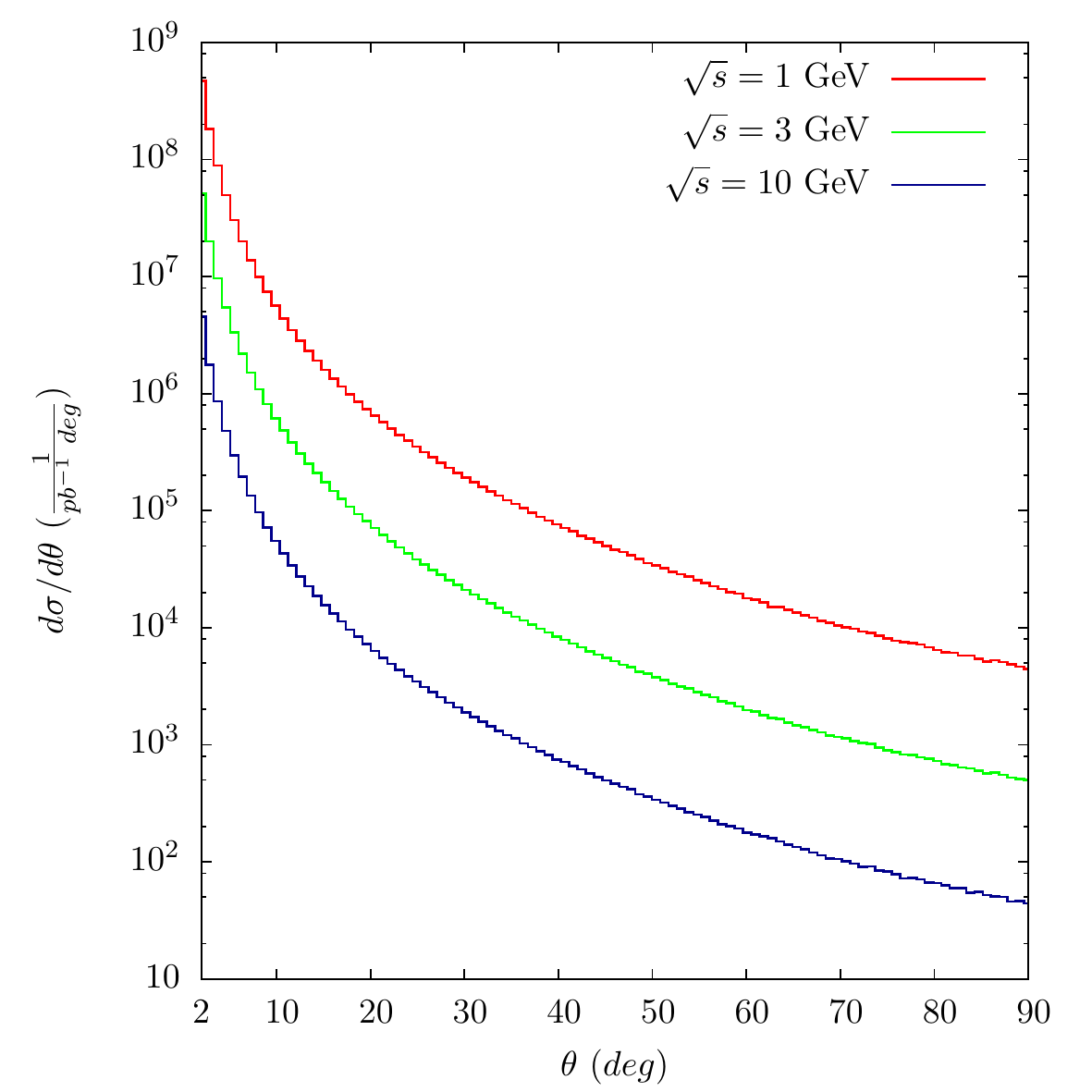}
  \caption{Left: Ranges of $x$ values as a function of the electron scattering angle $\theta$ for three different center-of-mass energies. The horizontal line corresponds to $x=x_{\rm peak}\simeq 0.914$. Right: Bhabha differential cross section obtained with \texttt{BabaYaga}~\cite{babayaga} as a function of $\theta$ for the same three values of $\sqrt s$ in the angular range $2^\circ<\theta<90^\circ$.}
  \label{xvsthnew}
\end{figure*}

Before entering the details of the extraction of $\Delta\alpha_{\rm had}(t)$ from Bhabha scattering data, let us consider a few simple points. In fig.~\ref{xfunction} (left) we plot the integrand $(1-x) \Delta \alpha_{\rm had} \! \left[ t(x) \right]$ of eq.~(\ref{amu_xalpha}) using the output of the routine \texttt{hadr5n12}~\cite{dalphaFred} (which uses time-like hadroproduction data and perturbative QCD). The range $x \in (0,1)$ corresponds to $t \in (-\infty,0)$, with $x=0$ for $t=0$. The peak of the integrand occurs at $x_{\rm peak}\simeq 0.914$ where $t_{\rm peak} \simeq-0.108$ GeV$^2$ and $\Delta\alpha_{\rm had}(t_{\rm peak}) \simeq 7.86 \times {10}^{-4}$ (see fig.~\ref{xfunction} (right)).
Such relatively low $t$ values can be explored at $e^+e^-$ colliders with center-of-mass energy $\sqrt s$ around or below 10 GeV (the so called ``flavor factories") where
\begin{equation}\label{kin}
t=-\frac{s}{2} \left(1-cos\theta \right) \left(1-\frac{4m_e^2}{s}\right),
\end{equation}
$\theta$ is the electron scattering angle and $m_e$ is the electron
mass. Depending on $s$ and $\theta$, the integrand of
eq.~(\ref{amu_xalpha}) can be measured in the range $x\in[x_{\rm
    min},x_{\rm max}]$, as shown in fig.~\ref{xvsthnew} (left). Note
that to span low $x$ intervals, larger $\theta$ ranges are needed as
the collider energy decreases. In this respect, $\sqrt s \sim 3$~GeV
appears to be very convenient, as an $x$ interval $[0.30,0.98]$ can be
measured varying $\theta$ between $\sim 2^\circ$ and $28^\circ$.
It is also worth remarking that data collected at flavor factories,
such as DA$\Phi$NE (Frascati), VEPP-2000 (Novosibirsk), BEPC-II
(Beijing), PEP-II (SLAC) and SuperKEKB (Tsukuba), and possibly at a
future high-energy $e^+e^-$ collider, like FCC-$ee$ (TLEP)~\cite{TLEP}
or ILC~\cite{ILC}, can help to cover different and complementary $x$ regions.

Furthermore, given the smoothness of the integrand, values outside the measured $x$ interval may be interpolated with some theoretical input. In particular, the region below $x_{\rm min}$ will provide a relatively small contribution to $a_{\mu}^{\scriptscriptstyle \rm HLO}$, while the region above $x_{\rm max}$ may be obtained by extrapolating the curve from $x_{\rm max}$ to $x=1$, where the integrand is null, or using perturbative QCD.

The analytic dependence of the MC Bhabha predictions on $\alpha(t)$ (and, in turn, on $\Delta\alpha_{\rm had}(t)$) is not trivial, and a numerical procedure has to be devised to extract it from the data.\footnote{This was not the case for example in~\cite{Arbuzov:2004wp,Abbiendi:2005rx}: there $\alpha(t)$ was extracted from Bhabha data in the very forward region at LEP, where the $t$ channel diagrams are by far dominant and $\alpha(t)$ factorizes.} 
In formulae, we have to find a function $\alpha(t)$ such that
\begin{equation}
\frac{d\sigma}{dt}\Big|_{\mathrm{data}} =
\frac{d\sigma}{dt}\Big(\alpha(t),\alpha(s)\Big)\Big|_{\mathrm{MC}},
\label{datavsMC1}
\end{equation}
where we explicitly kept apart the dependence on the time-like VP $\alpha(s)$ because we are only interested in $\alpha(t)$. We emphasise that, in our analysis, $\alpha(s)$ is an input parameter. Being the Bhabha cross section in the forward region dominated by the $t$-channel exchange diagram, we checked that the present $\alpha(s)$ uncertainty induces in this region a relative error on the $\theta$ distribution of less than $\sim 10^{-4}$ (which is part of the systematic error).

We propose to perform the numerical extraction of $\Delta\alpha_{\rm had}(t)$ from the Bhabha distribution of the $t$ Mandelstam variable. The idea is to let $\alpha(t)$ vary in the MC sample around a reference value and choose, bin by bin in the $t$ distribution, the value that minimizes the difference with data. The procedure can be sketched as follows:
\begin{enumerate}
\item choose a reference function returning the value of $\Delta\alpha_{\rm had}(t)$ (and hence $\alpha(t)$) to be used in the MC sample, we call it $\bar\alpha(t)$;
\item for each generated event, calculate $N$ MC weights by rescaling $\bar\alpha(t)\to\bar\alpha(t) + \frac{i}{N}\delta(t)$, where $i\in[-N,N]$ and $\delta(t)$ is for example the error induced on $\bar\alpha(t)$ by the error on $\Delta\alpha_{\rm had}(t)$. Being done on an event by event basis, the full dependence on $\alpha(t)$ of the MC
differential cross section can be kept;
\item for each bin $j$ of the $t$ distribution, compare the experimental differential cross section with the MC predictions and choose the $i_j$ which minimizes the  difference;
\item $\bar\alpha(t_j)+\frac{i_j}{N}\delta(t_j)$ will be the extracted value of $\alpha(t_j)$ from data in the $j^{th}$
  bin. $\Delta\alpha_{\rm had}(t_j)$ can then be obtained through  the relation between $\alpha(t)$ and $\Delta\alpha_{\rm had}(t)$.
\end{enumerate}
We finally find, for each bin $j$ of the $t$ distribution,
\begin{equation}
\frac{d\sigma}{dt}\Big|_{j,\mathrm{data}} =
\frac{d\sigma}{dt}\Big(\bar\alpha(t)+\frac{i_j}{N}\delta(t),\alpha(s)\Big)\Big|_{j,\mathrm{MC}}.
\label{datavsMC}
\end{equation}
We remark that the algorithm does not assume any simple dependence of the cross section on $\alpha(t)$, which can in fact be general, mixing $s$, $t$ channels and higher order radiative corrections, relevant (or not) in different $t$ domains.

In order to test our procedure, we perform a pseudo-experiment: we generate pseudo-data using the parameterization $\Delta\alpha_{\rm had}^I(t)$ of refs.~\cite{Hagiwara:2011af,dalphaTeub} and check
if we can recover it by inserting in the MC the (independent) parameterization $\Delta\alpha_{\rm had}^{II}(t)$ (corresponding to $\bar\alpha(t)$ of eq.~\ref{datavsMC}) of ref.~\cite{dalphaFred} by means of the method described above. For this exercise, we use the generator \texttt{BabaYaga} in its most complete setup, generating events at $\sqrt{s}=1.02$~GeV, requiring $10^\circ<\theta_\pm<170^\circ$, $E_\pm>0.4$~GeV and an acollinearity
cut of $15^\circ$. We choose $\delta(t)$ to be the error induced on $\alpha(t)$ by the $1$-$\sigma$ error on
$\Delta\alpha_{\rm had}(t)$, which is returned by the routine of ref.~\cite{dalphaFred}, we set $N=150$, and we produce distributions with 200 bins. We note that in the present exercise $\alpha(s)$ and all the radiative
corrections both in the pseudo-data and in the MC samples are exactly the same, because we are interested in testing the algorithm rather than assessing the achievable accuracy, at least at this stage.

\begin{figure*}[htp]
\centering
\includegraphics[width=.55\textwidth]{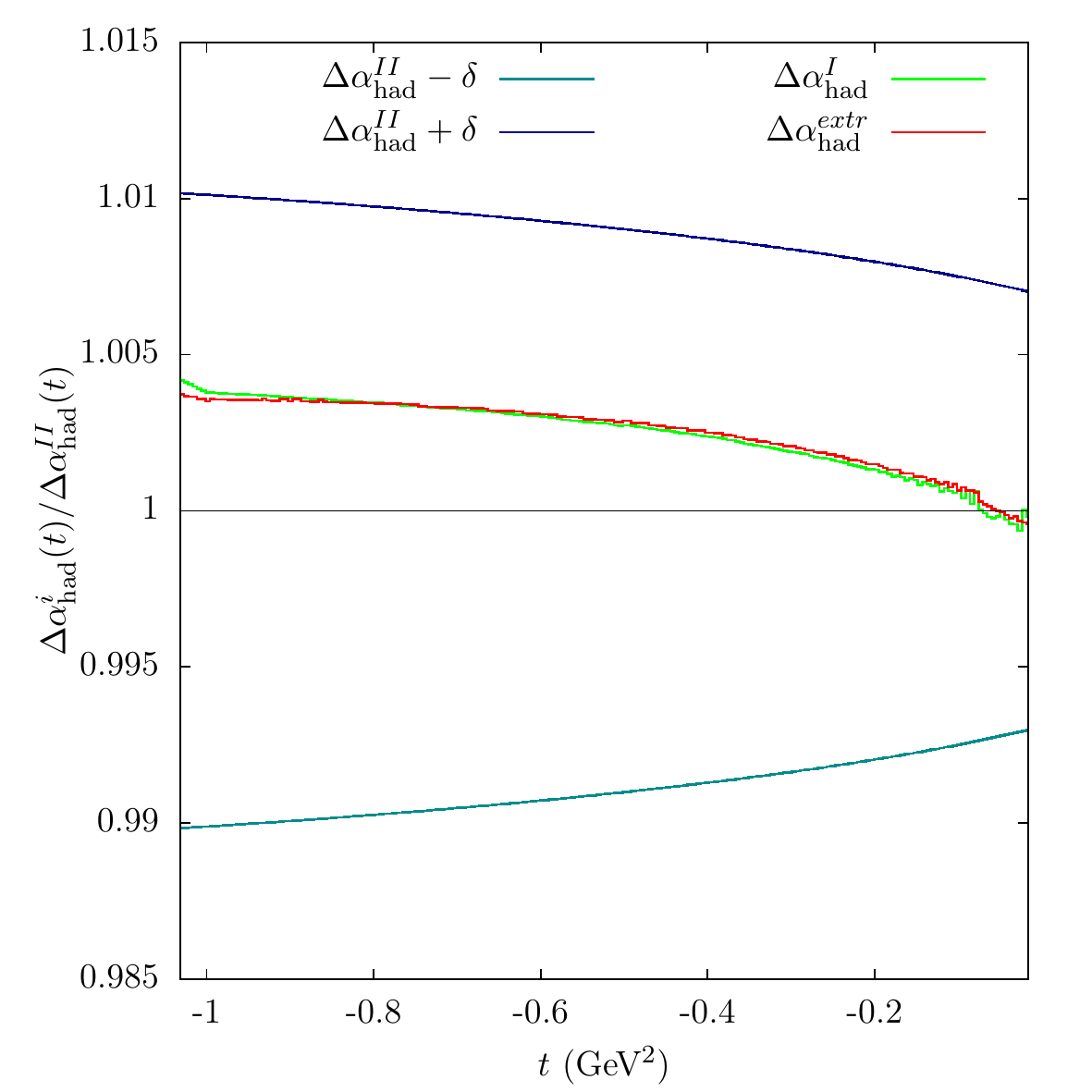}
  \caption{
The extracted function $\Delta\alpha_{\rm had}^{\rm extr}(t)$ compared to the function $\Delta\alpha_{\rm had}^{I}(t)$ used in the pseudo-data (see text). The functions $\Delta\alpha_{\rm had}^{II}(t) \pm \delta(t)$ are shown to display the range spanned by the MC samples. All functions have been divided by $\Delta\alpha_{\rm had}^{II}(t)$. The tiny difference between $\Delta\alpha_{\rm had}^{I}$ and $\Delta\alpha_{\rm had}^{\rm extr}$ is due to the binning discretization.}
  \label{extracted-ratio}
\end{figure*}

In fig.~\ref{extracted-ratio}, $\Delta\alpha_{\rm had}^{\rm extr}$ is the result extracted with our algorithm, corresponding to the minimizing set of $i_j$: the figure shows that our method is capable of recovering the underlying function $\Delta\alpha_{\rm had}(t)$ inserted into the ``data''. As the difference between $\Delta\alpha_{\rm had}^{I}$ and $\Delta\alpha_{\rm had}^{\rm extr}$ is hardly visible on an absolute scale, in fig.~\ref{extracted-ratio} all the functions have been divided by $\Delta\alpha_{\rm had}^{II}$ to display better the comparison between $\Delta\alpha_{\rm had}^{I}$ and $\Delta\alpha_{\rm had}^{\rm extr}$.

In order to assess the achievable accuracy on $\Delta\alpha_{\rm had}(t)$ with the proposed method, we remark that the LO contribution to the cross section is quadratic in $\alpha(t)$, thus we have (see eq.~(\ref{alphaq2}))
\begin{equation}
\frac{1}{2}\frac{\delta\sigma}{\sigma} \; \simeq \; \frac{\delta\alpha}{\alpha}
\;  \simeq \; \delta\Delta\alpha_{\rm had}
\label{dDalphahad}
\end{equation}
Equation (\ref{dDalphahad}) relates the {\it absolute} error on $\Delta\alpha_{\rm had}$ with the {\it relative} error on the Bhabha cross section. From the theoretical point of view, the present accuracy of the MC predictions~\cite{Actis:2010gg} is at the level of about $0.5\permil$, which implies that the precision that our method can, at best, set on $\Delta\alpha_{\rm had}(t)$ is $\delta\Delta\alpha_{\rm had}(t)\simeq 2\cdot 10^{-4}$. Any further improvement requires the inclusion of the NNLO QED corrections into the MC codes, which are at present not available (although not out of reach)~\cite{Actis:2010gg}.

From the experimental point of view, we remark that a measurement of
$a_{\mu}^{\scriptscriptstyle \rm HLO}$ from space-like data
competitive with the current time-like evaluations would require an
${\cal O} (1\%)$ accuracy. Statistical considerations show that a
$3\%$ fractional accuracy on the $a_{\mu}^{\scriptscriptstyle \rm
  HLO}$ integral can be obtained by sampling the integrand $(1-x)
\Delta \alpha_{\rm had} \! \left[ t(x) \right]$ in $\sim 10$ points
around the $x$ peak with a fractional accuracy of $10\%$. Given the
value of ${\cal O}(10^{-3})$ for $\Delta\alpha_{\rm had}$ at $x=x_{\rm
  peak}$, this implies that the cross section must be known with
relative accuracy of $\sim 2\times10^{-4}$. Such a statistical
accuracy, although challenging, can be obtained at flavor factories,
as shown in fig.~\ref{xvsthnew} (right). With an integrated luminosity
of ${\cal O}(1)$, ${\cal O}(10)$, ${\cal O}(100)$~$fb^{-1}$
at $\sqrt{s}=1$, $3$ and $10$ GeV, respectively,
the angular region of interest can be covered with a 0.01\% accuracy per degree. The experimental systematic error must match the same level of accuracy.

A fraction of the experimental systematic error comes from the knowledge of the machine luminosity, which is normalized by calculating a theoretical cross section in principle not depending on $\Delta\alpha_{\rm had}$. We devise two possible options for the normalization process:
\begin{enumerate}
\item using the $e^+e^-\to\gamma\gamma$ process, which has no dependence on $\Delta\alpha_{\rm had}$, at least up to NNLO order; 
\item using the Bhabha process at $t\sim 10^{-3}$~GeV$^2$ ($x \sim 0.3$), where the dependence on $\Delta\alpha_{\rm had}$ is of ${\cal O}(10^{-5})$ and can be safely neglected.
\end{enumerate}
Both processes have advantages and disadvantages; a dedicated study of the optimal choice goes beyond the scope of this paper and will be considered in a future detailed analysis.

\section{Conclusions}

We presented a novel approach to determine the leading hadronic
correction to the muon $g$-2 using measurements of the running of
$\alpha(t)$ in the space-like region from Bhabha scattering
data. Although challenging, we argued that this alternative
determination may become feasible
using data collected at present flavor
factories and possibly also at a future high-energy
$e^+e^-$ collider. The proposed determination can become
competitive with the accuracy of the present results obtained with the
dispersive approach via time-like data.

\section*{Acknowledgements}

\noindent We would like to thank G.~Degrassi, G.V.~Fedotovich,
F.~Jegerlehner and M.~Knecht for useful correspondence and
discussions. We would like also to thank G.~Montagna, F.~Piccinini and
O.~Nicrosini for constant interest in our work and useful
discussions. We acknowledge the hospitality of the Galileo Galilei
Institute in Florence, where part of this work has been carried out
during the workshop ``Prospects and Precision at the LHC at 14 TeV''.
C.M.C.C.~is fully supported by the MIUR-PRIN project 2010YJ2NYW. L.T. also acknowledges partial support from the same
MIUR-PRIN project.
M.P.~also thanks the Department of Physics and Astronomy of the
University of Padova for its support.
His work was supported in part by the MIUR-PRIN project 2010YJ2NYW and by the European Program INVISIBLES (PITN-GA-2011-289442).

\end{document}